\documentclass[twocolumn,a4paper,10pt,reprint,aip]{revtex4-1}

\usepackage{graphicx}
\usepackage{latexsym}   
\usepackage{amsmath,amssymb,amsthm}

\DeclareMathOperator{\sech}{sech}

\usepackage{pifont}
\usepackage{mathrsfs}
\usepackage{isomath}
\usepackage{xcolor}

\usepackage[hidelinks]{hyperref}
\hypersetup{
    colorlinks = true,
    citecolor = blue,
    linkcolor=blue
}

\makeatletter
\renewcommand{\p@subsection}{}
\renewcommand{\p@subsubsection}{}
\makeatother

\begin{document}

\title{Revisiting the strain-induced softening behaviour in hydrogels}

\author{L. K. R. Duarte}

\affiliation{Departamento~de~F\'isica,~Universidade~Federal~de~Vi\c{c}osa~(UFV),~36.570-900,~Vi\c{c}osa,~MG,~Brazil.}

\affiliation{Instituto Federal de Educa\c{c}\~ao,~Ci\^encia e Tecnologia de Minas Gerais,~35.588-000,~Arcos, MG, Brazil.}

\author{L. G. Rizzi}

\affiliation{Departamento~de~F\'isica,~Universidade~Federal~de~Vi\c{c}osa~(UFV),~36.570-900,~Vi\c{c}osa,~MG,~Brazil.}


\begin{abstract}
	Usually, the strain-induced softening behaviour observed in the differential modulus $K(T,\gamma)$ of hydrogels has been attributed to the breakage of internal structures of the network, such as the cross-links that bind together the polymer chains.
	Here we consider a stress-strain relationship that we have recently derived from a coarse-grained model to demonstrate that no rupture of the network is needed for rubber-like gels to present such behaviour.
	In particular, we show that, in some cases, the decreasing of $K(T,\gamma)$ as a function of the strain $\gamma$ is closely related to the energy-related contribution to the elastic modulus that has been experimentally observed, {\it e.g.}, for tetra-PEG hydrogels.
	Thus, our results suggest that, instead of the breakage of structures, the softening behaviour can be also related to the effective interaction between the chains in the network and their neighbouring solvent molecules.
	Comparison to experimental data determined for several hydrogels is included to illustrate that behaviour and to validate our approach.
%
\end{abstract}

\maketitle


\section{Introduction}
\label{intro}

\begin{figure}[!t]
	\centering
	\resizebox{0.44\textwidth}{!}{\includegraphics{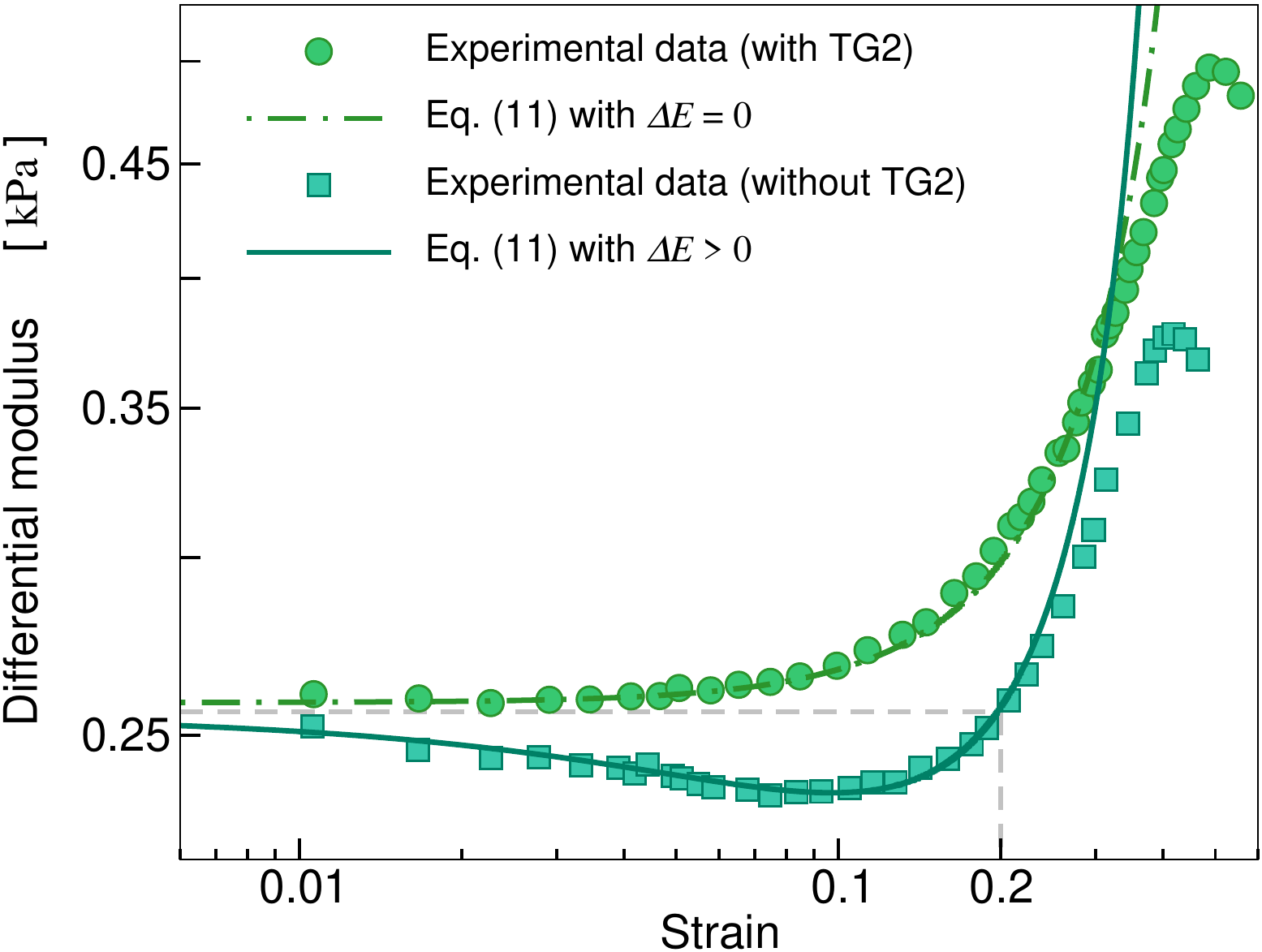}}
	\caption{Differential modulus $K(T,\gamma)$ as a function of the strain $\gamma$ for two different samples of a collagen-based hydrogel, {\it i.e.}, with and without crosslinking agent (transglutaminase II, TG2).
	Dash-dotted and continuous lines correspond to the results obtained from Eq.~\eqref{eqn:differential-modulus} with $\Delta E = 0$ and $\Delta E > 0$, respectively (see Sec.~\ref{sec:biologycal} for the values of all parameters).
	Filled symbols correspond to the experimental data extracted from Ref.~\cite{valero2018plosone}.
}
	\label{fig:collagen_intro}
\end{figure}

	Hydrogels, like many soft materials, can display a strain-induced softening behaviour which is characterized by a decrease in the differential modulus at intermediate deformations.
	Figure~\ref{fig:collagen_intro} illustrates such behaviour for a collagen-based hydrogel, 
with the differential modulus defined as
\begin{equation}
K(T,\gamma) = \left[\dfrac{\partial }{\partial \gamma}\sigma(T,\gamma)\right]_T~~,
\label{eqn:differential-modulus_def}
\end{equation}
where $\sigma(T,\gamma)$ is the stress for a given strain $\gamma$ at a given temperature $T$.
	Investigating the origin of such effect in hydrogels is crucial for gaining insights into a wide range of viscoelastic materials~\cite{ferraro2023softmatter,richtering2014sm}, which includes not only biopolymer-based gels~\cite{hyungsuk2010physrev,orakdogen2010macromo,kurniawan2012biomacromol,kraxner2021nanoscale}, but also industry-related gels that are related to innovative therapies and medical applications~\cite{annabi2014advmater,thiele2014advmater,kamata2015advhmater,calo2015eurpolj,basu2011macromol,wen2012softmatter,dai2021softmatter}.

	In general, the softening phenomenon is attributed either to the breakage of internal structures of the material, as in the case of reversibly crosslinking in peptide gels~\cite{greenfield2010langmuir} and polymer networks~\cite{xucraig2011macromolecules,burla2019natrevs}, or to the yielding, as in the case of soft glasses~\cite{rogers2019rheolacta}.
	Usually, the differential modulus $K(T,\gamma)$ defined by Eq.~\eqref{eqn:differential-modulus_def} is evaluated from tensile experiments, but it is worth noting that it can be also estimated from oscillatory experiments~\cite{larsonbook}, where in the later one has that $K(T,\gamma) \approx \lim_{\omega \rightarrow 0} G^\prime(\omega,T,\gamma)$, with $G^\prime$ being the storage modulus. 
	Hence, one also has that $G(T) = \lim_{\gamma \rightarrow 0}K(T,\gamma)$ corresponds to the usual elastic ({\it i.e.}, shear) modulus that characterizes the mechanical response of isotropic semisolid materials at their linear viscoelastic (LVE) regime.
	Oscillatory experiments can be particularly important since they can easily identify the breakage of internal structures of the gel, which may become a liquid-like material at high enough strains, so the loss modulus $G^{\prime \prime}$ surpasses the storage modulus $G^{\prime}$ (see, {\it e.g.}, Refs.~\cite{greenfield2010langmuir,rogers2019rheolacta,yu2021natcommun}).

	Intriguingly, some hydrogels do display a strain-induced softening behaviour as the one illustrated by Fig.~\ref{fig:collagen_intro} while keeping their semisolid character unchanged, {\it i.e.}, with $G^{\prime} > G^{\prime \prime}$.
	Examples include not only the collagen-based hydrogel of Ref.~\cite{valero2018plosone}, but also several other gels, including those based on tetra-PEG polymers~\cite{kamata2014science,kamata2015advhmater} and some based on biofilaments~\cite{storm2005nature,motte2012biopolymers,oztoprak2017intjbiomol}.
	Indeed, numerical simulations of sparsely connected networks~\cite{bouzid2018langmuir} indicate that the softening behaviour observed in the differential modulus can occur even without breaking the bonds of the gel network.
	The caveat here is that the decrease in $K$ observed in Ref.~\cite{bouzid2018langmuir} is obtained from simulations where only enthalpic contributions were considered, which is the usual approach assumed in many numerical studies of semiflexible polymer networks~\cite{heidenmann2015softmatter,huisman2010pre,huisman2011prl}.
	Given the significant presence of solvent within the network structure of the hydrogels, the assumption of 
an athermal~\cite{rizzi2022fip}, {\it i.e.}, purely enthalpic, mechanical response is in clear contrast to the purely entropic approaches, such as the ones considered in the classical theories~\cite{flory1953book,james1953jcp,flory1977jcp}.
	Recent experimental~\cite{yoshikawa2021prx,sakumichi2021polymj,aoyama2023ascmacro} and computational studies~\cite{nobu2023phys,hagita2023macromo} have indicated, however, that purely entropic, {\it e.g.}, rubber-based~\cite{treloarbook}, theories are also insufficient for describing the mechanical properties of hydrogels, and a significant energy-related contribution is required to describe their elastic behaviour.

	Although models based on worm-like chains (WLC) are expected to encompass the stress-strain relationship of biopolymer hydrogels~\cite{palmer2008acta,mackintosh1995physrev}, in most cases they seem to be able to describe only their hardening behaviour.
	In principle, the lack of the softening effect can be rationalized in terms of the well-known force-extension relation introduced by Marko \& Siggia~\cite{marko1995macromo}, where the force $f_{\text{WLC}}$ is a monotonically increasing function of the deformation, with $\partial f_{\text{WLC}}/\partial z >0$
and $\partial^2 f_{\text{WLC}}/\partial z^2 >0$ for any value of the deformation $z$.
	Thus, the same strictly increasing behaviour is expected for its differential modulus, 
and no softening behaviour occurs since $K_{\text{WLC}}$ is proportional to the first derivative of $f_{\text{WLC}}$.
	Alternatively, one can access the response of hydrogels due to both enthalpic and entropic contributions by considering  a simple coarse-grained model that we have recently introduced in Ref.~\cite{duarte2023epje}, which takes into account the interaction of the chains in the network with their neighbouring solvent molecules.
	Hence, the main idea of this study is to consider the possibility that such model could explain the strain-induced softening behaviour that is being observed in several hydrogels, including the one related to the results displayed in Fig.~\ref{fig:collagen_intro}.

	The remainder of the paper is given as follows.
	Section 2 include the minimum information about the model presented in Ref.~\cite{duarte2023epje}, and also the relevant information about its connection to the experimental results and the phenomenological model of Ref.~\cite{yoshikawa2021prx}.
	In Sec. 3 we derive the theoretical expression for the differential modulus $K(T,\gamma)$ and other relationships between experimentally relevant quantities.
	In that section we also present the main results and validate our approach through comparisons to experimental data obtained from different hydrogels, including several synthetic and biopolymer-based ones.
	Section 4 include some concluding remarks where we discuss the relevance and limitations of our approach.

\section{Previous results}

\subsection{Theoretical results from Duarte \& Rizzi~\cite{duarte2023epje}}
\label{duarterizzimodel}

As demonstrated in Ref.~\cite{duarte2023epje}, rubber-like hydrogels can be described by a stress-strain relationship that is given by
\begin{equation}
\sigma(T,\gamma) = \bar{n}\left\{ \Delta E + k_BT \ln \left[ \frac{\gamma_{b}(T) + \gamma}{ \gamma_{s}(T) - \gamma } \right] \right\}~,
\label{eqn:sigma-gamma-q}
\end{equation}
where $\bar{n} =n_e(q_{\ell}^{\,}+1)/[2(q_{\ell}^{\,}-1)]$, with $n_{e}$ being related to the number density of elastic elements within the network, {\it i.e.}, chains which are connected by cross-links, and $q_{\ell}= - \ell_s/\ell_b$.
	Equation~\eqref{eqn:sigma-gamma-q} is obtained from a two-state model, so that the parameters $\ell_{b}$ and $\ell_{s}>|\ell_{b}|$ correspond to the end-to-end displacements of the blob ({\it b}) and stretched ({\it s}) conformational states which are available for each of the $N$ segments that compose a chain that connects two cross-links in the gel network (see Ref.~\cite{duarte2023epje} for further details).
	It is assumed that these two possible states have energies $E_b$ and $E_s$, so that $\Delta E=E_s-E_b$ corresponds to the energy difference between the $s$ and $b$ conformations.
	The temperature-dependent functions  $\gamma_s(T)$ and $\gamma_b(T)$ should also depend on the values of $q_{\ell}$ and $\Delta E$, thus making Eq.~\eqref{eqn:sigma-gamma-q} with just three parameters.
	Explicitly, these functions can be written as~\cite{duarte2023epje}
\begin{equation}
\gamma_{b}(T)=\frac{q_{\ell}^{\,}+1}{q_{\ell}^{\,}-e^{\Delta E/k_BT}} ~,
\label{eqn:gammab-q}
\end{equation}
and 
\begin{equation}
\gamma_{s}(T)  = \frac{q_{\ell}^{\,}+1}{q_{\ell}^{\,}e^{-\Delta E/k_BT}-1}~.
\label{eqn:gammas-q}
\end{equation}

	Each chain in the gel network may be stretched up to a maximum length $\ell_c=N\ell_s$, which is approached when the applied stress becomes large enough.
	In addition, the natural end-to-end distance of a chain in the gel network is determined in the absence of stress and is given by~\cite{duarte2023epje}
\begin{equation}
\ell_{0}^{\,}(T) = \dfrac{\ell_c}{q_{\ell}^{\,}}\left(\dfrac{q_{\ell}^{\,}-e^{\Delta E/k_BT}}{1+e^{\Delta E/k_BT}}\right)~.
\label{eqn:ell_0}
\end{equation}
	The above expression indicates that $\ell_{0}^{\,}=\ell_{0}^{\,}(T)$ is a fraction of its maximum contour length $\ell_c$, regardless if $q_{\ell}^{\,}$ is positive or negative.
	With $|q_{\ell}|>1$, the distance $\ell_{0}^{\,}>0$ allows one to define the strain $\gamma=(\ell-\ell_{0}^{\,})/\ell_{0}^{\,}$, with $\ell>0$ denoting the mean end-to-end distance between cross-links when stress is applied to the network~\cite{duarte2023epje}.
	Moreover, one may infer that $\gamma_s$, as defined by Eq.~\eqref{eqn:gammas-q}, is the maximum strain supported by a chain in the network.
	In fact, if the integrity of the internal structures of the network are to be preserved, 
$\gamma_s$ also corresponds to the maximum strain supported by the hydrogel as a whole.
	Its role as the maximum strain is also confirmed by Eq.~\eqref{eqn:ell_0}, from where one can find that $\gamma_s = (\ell_c-\ell_{0}^{\,})/\ell_{0}^{\,}$, which is precisely the result one would get from expression~\eqref{eqn:gammas-q}.

	From the stress-strain relationship, Eq.~\eqref{eqn:sigma-gamma-q}, one can obtain the elastic modulus of the hydrogel through its definition, that is, 
\begin{equation}
G(T)
= \lim_{\gamma \rightarrow 0}\left[ \frac{ \partial}{ \partial \gamma} \sigma(T,\gamma) \right]_T 
=  n_ek_BT\phi(T)~,
\label{eqn:elastic-modulus-w}
\end{equation}
where the temperature-dependent factor is given by
\begin{equation}
\phi(T) = \left(\dfrac{1+e^{\Delta E/k_BT}}{2}\right)\left(\frac{q_{\ell}^{\,}e^{-\Delta E/k_BT} - 1}{q_{\ell}^{\,}-1}\right)~.
\label{eqn:wT-q}
\end{equation}
	It is worth noting that, if there is no energy difference between the $s$ and $b$ conformational states, {\it i.e.}, $\Delta E=0$, then $\phi(T)=1$, at any temperature.
	In this case, the model recovers the usual result obtained from purely entropic cross-linked polymer networks~\cite{rubinsteinbook}, where the elastic modulus is equal to its positive contribution, {\it i.e.}, $G(T) = G_S(T) = n_ek_BT$.
	Hence, by assuming that the elastic modulus is given by $G(T)=G_S(T)+G_E(T)$ as in Refs.~\cite{yoshikawa2021prx,sakumichi2021polymj}, one have that the energy-related contribution $G_E(T)$ is zero.

	On the other hand, when $\Delta E >0$, the function $\phi(T)$ behaves differently when describing distinct types of hydrogels.
	For $q_{\ell}^{\,}>1$, Eq.~\eqref{eqn:wT-q} revealed the existence of a temperature $T_0^*$, which defines a range of temperatures where the elastic modulus $G(T)$ is positive, that is, $T>T_0^*$.
	In this case, the energy-related contribution to the elastic modulus is no longer zero, and, in fact, it is significantly negative, $G_E(T)<0$.
	As discussed in Ref.~\cite{duarte2023epje}, for $q_{\ell}^{\,}>1$, the temperature $T_0^*$ is found to be given by
\begin{equation}
T_0^* = \dfrac{\Delta E}{k_B\ln q_{\ell}^{\,}}~.
\label{eqn:T_0*}
\end{equation}
When $q_{\ell}^{\,}<-1$, however, one finds from Eq.~\eqref{eqn:wT-q} that no threshold temperature exists, since $\phi(T)$ is positive for any temperature. 
	In this case, the energy-related contribution $G_E(T)$ may be either positive or negative, depending on the temperature of the system.
	This later case, where $q_{\ell}^{\,}<-1$, will be discussed in detail in Sec.~\ref{sec:biologycal}, while the former case, where $q_{\ell}^{\,}>1$, is directly related to the experiments and to the phenomenological model presented in Ref.~\cite{yoshikawa2021prx}, which are summarized in the following.

\subsection{Connections to the results from Yoshikawa et al.~\cite{yoshikawa2021prx}}
\label{sec:negative-contribution}

	In Ref.~\cite{yoshikawa2021prx}, Yoshikawa \textit{et. al.} presented extensive experimental evidence for a negative energy-related contribution to the elastic modulus for tetra-PEG hydrogels, which are synthetically designed gels with well established topology.
	In order to describe their data, they introduced a phenomenological model where the stress-strain relationship is expressed as~\cite{yoshikawa2021prx}
\begin{equation}
\sigma(T,\gamma) = a(T-T_0)\,\gamma~~,
\label{eqn:stress-strain-adhoc}
\end{equation}
with $a$ and $T_0$ being system-dependent positive phenomenological constants.
	The above expression indicates that the elastic modulus is simply given by $G(T)=a(T-T_0)$.
	Thus, by writing it as $G(T)=G_S(T) + G_E(T)$, the authors of Ref.~\cite{yoshikawa2021prx} attributed the term $G_S(T) =aT$ to a positive ``entropic'' contribution, while the negative one, $G_E(T)=G(T)-G_S(T)=-aT_0$, is assumed to be an ``energetic'' (or ``enthalpic'') contribution to the elastic modulus
	(here we refer the reader to the Appendix of Ref.~\cite{duarte2023epje} for a detailed discussion about the correct meaning of the claimed ``entropic'' and ``enthalpic'' contributions).

	Although the mechanical properties of tetra-PEG hydrogels were somewhat well described in Refs.~\cite{yoshikawa2021prx,sakumichi2021polymj}, the theoretical basis for the phenomenological expression~\eqref{eqn:stress-strain-adhoc} was not clear
until we showed that it is a limiting case of the expression obtained from the model we proposed in Ref.~\cite{duarte2023epje}.
 	In particular, by considering Eqs.~\eqref{eqn:elastic-modulus-w} and~\eqref{eqn:wT-q} one can deduce the following expression for the energy-related contribution to the elastic modulus
\begin{equation}
G_E(T) =-\dfrac{\Delta E}{k_BT}\,\dfrac{q_{\ell}^{\,}+e^{2\Delta E/k_BT}}{(q_{\ell}^{\,}-e^{\Delta E/k_BT})(1+e^{\Delta E/k_BT})}\,G(T)~.
\label{eqn:G_E-G-w}
\end{equation}
	Indeed, as one may check, by considering $e^x\approx 1+x$ and $q_{\ell}>1$, the above expression is approximately equal to the simpler phenomenological form, {\it i.e.}, $G_E(T)=-a T_0$, 
with the temperature $T_0$ associated to the temperature $T_0^{*}$  given by Eq.~\eqref{eqn:T_0*}.
	Similarly, one can show that the above expression obtained from our model leads to a $G_S(T)$ contribution which can be approximated to a term that is proportional to the temperature~\cite{duarte2023epje}, just as in the phenomenological model, {\it i.e.}, $G_S(T)=aT$.

	Importantly, the experiments done in Refs.~\cite{yoshikawa2021prx,sakumichi2021polymj} revealed that the negative contribution to the elastic modulus can be as high as $|G_E(T)|\approx 2G(T)$, so that $0<G(T) < G_S(T)$, highlighting the significance of this energy-related contribution to the elastic modulus $G(T)$.
	We note that Eq.~\eqref{eqn:G_E-G-w} is valid for both $q_{\ell}^{\,}>1$ and $q_{\ell}^{\,}<-1$, however, since both $G(T)$ and $T_0^*$ are assumed to be positive constants for tetra-PEG hydrogels, one can argue that $G_E(T)$  will be negative and $T_0^*$ (Eq.~\eqref{eqn:T_0*}) will be positive only if $\Delta E>0$ and $q_{\ell}>1$.
	Indeed, these were the ranges we have considered for those parameters in Ref.~\cite{duarte2023epje} in order to describe the whole set of experimental data obtained for tetra-PEG hydrogels presented in Refs.~\cite{yoshikawa2021prx,sakumichi2021polymj}.
	It is worth recalling that $\Delta E$ and $q_{\ell}$ are, respectively, the energy difference and the ratio between distances for the two possible conformational states, $s$ and $b$, of a segment of a chain that connects cross-links in the gel network.
	By considering positive values for these two parameters, one is assuming not only that the blob state has a smaller energy, $E_b < E_s$, but also that $\ell_b$ is negative, which mean that in such state the segment of the chain is folded around itself (see the schematic drawing in Ref.~\cite{duarte2023epje} for a clearer picture).
	While the positive value of $\Delta E$ is supported by many numerical simulations which assume effective interactions between the chains and the solvent~\cite{zierenbergreview}, the negative value of $\ell_{b}$ can be interpreted as a somewhat non-affine behaviour~\cite{boycearrudareview} of the segments within the chains in the gel network, since $b$ states tend to decrease the distance between cross-links at the mesoscopic scale.
	Indeed, such non-affine behaviour can be observed from experiments and numerical simulations of single PEG chains~\cite{kolberg2019JACS} and it can be clearly related to the effective interaction between the polymer chain and the solvent molecules, which corroborated our approach.
	Even so, one should note that our model does not require that $\ell_b<0$ and $q_{\ell}>1$. 
	In fact, we will show later in this study that some hydrogels can be only described by Eq.~\eqref{eqn:sigma-gamma-q} if $\ell_b>0$ and $q_{\ell}<-1$.

	Finally, it is worth emphasizing that, despite the agreement between the phenomenological model and our approach as a limiting case, if the model of Ref.~\cite{yoshikawa2021prx} is considered, the stress-strain relationship given by Eq.~\eqref{eqn:stress-strain-adhoc} leads to a strain-independent differential modulus, that is, $K(T,\gamma) = a(T-T_0^*)$.
	Hence, such phenomenological model is not able to describe the strain-induced softening behaviour illustrated in Fig.~\ref{fig:collagen_intro}.
	In the following we show that is not the case of the stress-strain relationship obtained from our model, previously discussed in Sec.~\ref{duarterizzimodel}.

\section{Results}
\label{sec:results}

	Clearly, as we have discussed above, the WLC-based models~\cite{palmer2008acta} and the phenomenological model of Ref.~\cite{yoshikawa2021prx} are incapable of describing the strain-induced softening behaviour observed for hydrogels like the one illustrated by the data displayed in Fig.~\ref{fig:collagen_intro}. 
	In order to investigate that issue, we now consider the stress-strain relationship given by Eq.~\eqref{eqn:sigma-gamma-q} and evaluate the differential modulus through its definition, {\it i.e.}, Eq.~\eqref{eqn:differential-modulus_def}, which yields
\begin{equation}
K(T,\gamma) = \dfrac{ G(T)}{\left[1+\gamma/\gamma_b(T)\right]\left[1-\gamma/\gamma_s(T)\right]}.
\label{eqn:differential-modulus}
\end{equation}
	Here, $G(T)$ is the elastic modulus given by Eq.~\eqref{eqn:elastic-modulus-w}, and  $\gamma_b(T)$ and $\gamma_s(T)$ are defined as in Eqs.~\eqref{eqn:gammab-q} and \eqref{eqn:gammas-q}, respectively. 
	Accordingly, as $\gamma$ approaches zero, the above expression indicates that the differential modulus will tend to the elastic modulus, {\it i.e.}, $G(T)=\lim_{\gamma \rightarrow 0}K(T,\gamma)$.
	On the other hand, when the strain approaches $\gamma_s(T)$, the differential modulus will increase, indicating that the hydrogel is becoming harder.

	Below we first discuss what happens for intermediate values of the strain and relate the mechanical response of the hydrogel with the non-linear strain-induced softening behaviour similar to what is shown in Fig.~\ref{fig:collagen_intro}. 
	Besides comparing our theoretical results to the data extracted from experiments to validate our approach, we also include a discussion about the hardening behaviour in view of known results presented in the literature.

\subsection{Strain-induced softening behaviour}
\label{sec:softening}

	Importantly, from Eq.~\eqref{eqn:differential-modulus} one observes that the value of the differential modulus will reach a minimum for finite values of $\gamma$, where the denominator of $K(T,\gamma)$ reaches a maximum.
	This corresponds to a value of strain that is given by 
\begin{equation}
\gamma_\textrm{min}(T) =\gamma_s(T)\left(\dfrac{1-e^{-\Delta E/k_BT}}{2}\right)~.
\label{eqn:gamma-min}
\end{equation}
	At this strain, the minimum value of the differential modulus is given by
\begin{equation}
K_{\min}(T) = G(T)\,\text{sech}^2 \left(\dfrac{\Delta E}{2k_BT}\right).
\label{eqn:K-min}
\end{equation}

	In addition, by replacing the value obtained for $\gamma_{\min}(T)$ in the stress-strain relationship given by Eq.~\eqref{eqn:sigma-gamma-q}, one finds the corresponding stress
\begin{equation}
\sigma_{\min}= \bar{n}\Delta E.
\label{eqn:sig_min}
\end{equation}
	Interestingly, the above result indicates that the stress which minimizes the differential modulus is independent of the temperature.
	In fact, $\sigma_{\min}$ will only depend on the number density $\bar{n}$ (which, in turn, depends on $n_e$ and $q_{\ell}^{\,}$), and on the energy difference $\Delta E$. 
	Obviously, all the mentioned parameters are characteristic to the hydrogel itself (see Ref.~\cite{duarte2023epje} for further details).

\subsubsection{Case $\Delta E =0$}
\label{sec:Eeq0}

If $\Delta E=0$, then $\textrm{sech}(0)=1$, thus Eq.~\eqref{eqn:K-min} indicates that the minimum value of the differential modulus is just $K_\textrm{min}(T)=G(T)$. 
	In that case, Eq.~\eqref{eqn:differential-modulus} becomes a strictly increasing function of $\gamma$ which is given by $K(T,\gamma) =G(T) \{1-[\gamma/\gamma_{s}(T)]^2\}^{-1}$ (since $\gamma_{b}(T)=\gamma_{s}(T)$, as established by Eqs.~\eqref{eqn:gammab-q} and \eqref{eqn:gammas-q}). 
	This implies that $K(T,\gamma)\ge G(T)$ for any strain in the range $0<\gamma <\gamma_{s}(T)$, and at any temperature $T>0$, since from Eq.~\eqref{eqn:T_0*} one has that $T_0^* = 0$ when $\Delta E=0$.
	Besides, Eqs.~\eqref{eqn:gamma-min} and \eqref{eqn:sig_min} indicate that $\gamma_{\min}(T)=0$ and $\sigma_{\min}=0$, consequently, our model predicts that there is no strain-induced softening behaviour in the cases where $\Delta E=0$.
	As it may be checked, this also corresponds to gel networks with a purely entropic response, because, when the energy difference $\Delta E$ is zero, it follows from Eq.~\ref{eqn:G_E-G-w} that $G_E(T)=0$, thus $G(T)=G_S(T)$.

	Therefore, for hydrogels where there is no energy difference, one can only observe the hardening behaviour for any value of the strain, {\it i.e.},  $0<\gamma \lesssim \gamma_s(T)$.
	This is precisely what is shown in Fig.~\ref{fig:collagen_intro} for the sample of a collagen-based hydrogel with TG2 crosslinking agent, where the differential modulus displays a strictly increasing behaviour up to relatively large strains, {\it i.e.}, $\gamma < 0.45$.
	Clearly, for even higher values of the strain, {\it i.e.}, $\gamma \gtrsim 0.45$,
the experimental data displayed in Fig.~\ref{fig:collagen_intro} indicate that there is a softening behaviour for both samples, but we argue here that such behaviour is, in fact, due to the breakage of the internal structures of the gel network, so it is beyond the scope of our model.

\begin{figure*}[!t]
	\centering
	\resizebox{0.74\textwidth}{!}{\includegraphics{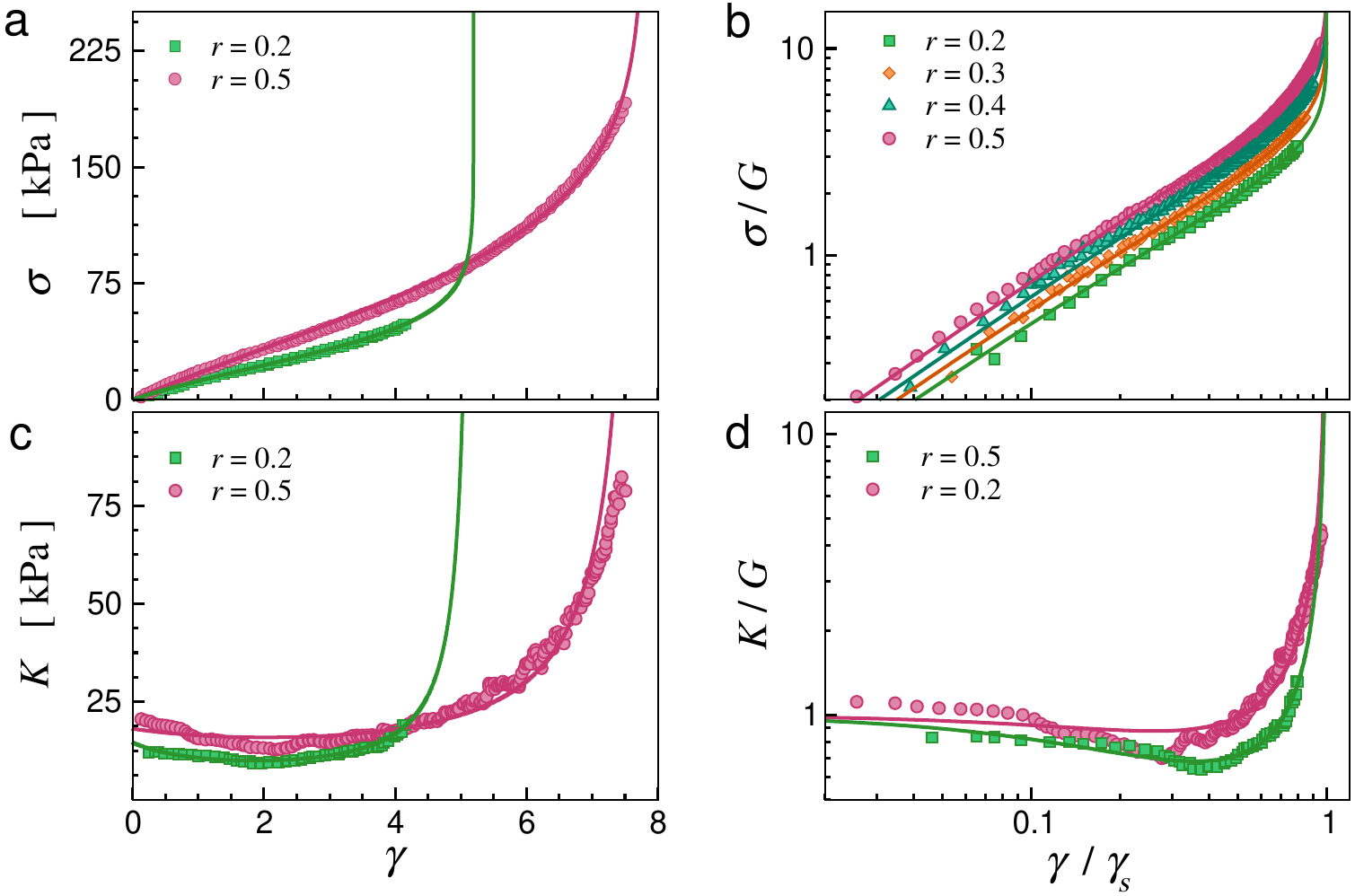}}
	\caption{Top panels display the stress-strain curves:
(a)~$\sigma\times \gamma$ in linear scale and 
(b)~$\sigma/G\times\gamma/\gamma_s$ in log-log scale;
	while bottom panels display the	differential modulus:
(c)~$K\times\gamma$ in linear scale and 
(d)~$K/G\times\gamma/\gamma_s$ in log-log scale. 
	Filled symbols correspond to the experimental data extracted from Ref.~\cite{kamata2014science}, where PEG hydrogels with different hydrophobic ratios $r$ were studied.
	Continuous lines correspond to the stress-strain and differential modulus expressions given by Eqs.~\eqref{eqn:sigma-gamma-q}~and~\eqref{eqn:differential-modulus}, respectively, evaluated with the parameters listed in Table~\ref{tab:kamatta_parameters}.}
	\label{fig:kamatta}
\end{figure*}

\subsubsection{Case $\Delta E > 0$}

	Because $\sech(x)\le 1$ for $x \neq 0$, Eq.~\eqref{eqn:K-min} establishes that, if $\Delta E\ne 0$, the minimum value of the differential modulus will correspond to a fraction of the elastic modulus, with $K_\textrm{min}(T)<G(T)$. 
	Besides, Eq.~\eqref{eqn:differential-modulus} indicates that $K(T,\gamma)$ will decrease for $0<\gamma<\gamma_\textrm{min}(T)$, reaching its minimum value when $\gamma=\gamma_\textrm{min}(T)$. 
	From this value up to $\gamma_{s}(T)$, one observes that the differential modulus increases indefinitely, indicating then a hardening behaviour. 
	Interestingly, it can be shown from Eq.~\eqref{eqn:differential-modulus} that $G(T)=K(T,2\gamma_\textrm{min}(T))$, which is an identity that may be used to better determine the value of $\gamma_\textrm{min}(T)$ in the cases where $K_\textrm{min}(T)$ cannot be clearly identified from the experimental data.
	For $\gamma < 2\gamma_\text{min}(T)$ the gel is softer than before being deformed ({\it i.e.}, $\gamma=0$), and, when $\gamma > 2\gamma_\text{min}(T)$, it is harder.
	So we consider that the strain-induced softening behaviour happens for $0<\gamma <\gamma_{\min}(T)$, while the hardening behaviour occurs in the range $\gamma_{\min}(T)< \gamma \lesssim\gamma_{s}(T)$. 
	This kind of response is illustrated by the data presented in Fig.~\ref{fig:collagen_intro} for the collagen-based hydrogel without TG2 crosslinking agent.
	One can easily identify $\gamma_{\min}(T)=0.1$ since, at $\gamma=0.2=2\gamma_{\min}(T)$, the differential modulus is equal to the elastic modulus, {\it i.e.}, $G(T)=K(T,2\gamma_{\min}(T))\approx 0.25\,$kPa (dashed horizontal line).
	Thus, we assume there that the strain-induced softening behaviour occurs for $\gamma<0.1$, whereas the hardening behaviour is observed for $\gamma > 0.1$ (obviously excluding the range of even higher strains, {\it i.e.}, $\gamma \gtrsim 0.45$, as indicated in the last subsection).

\begin{table*}
\small
  \caption{Here, $n_e$, $q_\ell^{\,}$ and $\Delta E$ are parameters obtained from a fit procedure of Eq.~\eqref{eqn:sigma-gamma-q} to the experimental data extracted from Ref.~\cite{kamata2014science}, where the stress-strain of samples of PEG hydrogels with different hydrophobic ratios $r$ are presented.
	The values of $\gamma_{s}$, $G$, $\gamma_\textrm{min}$, and $K_\textrm{min}$ were obtained from these three parameters through Eqs.~\eqref{eqn:gammas-q},~\eqref{eqn:G_E-G-w},~\eqref{eqn:gamma-min}, and~\eqref{eqn:K-min}, respectively.
	Expressions~\eqref{eqn:T_0*} and~\eqref{eqn:G_E-G-w} were used to estimate the temperature $T_0^*$ and the ratio $G_E/G$, respectively.}
  \label{tab:kamatta_parameters}
  \begin{tabular*}{\textwidth}{@{\extracolsep{\fill}}cccccccccc}
    \hline
$r$ & $n_e\,$[nm$^{-3}$] & $q_\ell^{\,}$ & $\Delta E\,$[pN.nm] & $\gamma_s$ & $G\,$[kPa] & $\gamma_\textrm{min}$ & $K_\textrm{min}\,$[kPa] & $G_E/G$ & $T_0^*\,$[K]  \\
    \hline
		$0.2$ & $0.00666$ & $13.50$ & $5.44$ & $5.198$ & $14.512$ & $1.87$ & $9.934$ & $-0.73$ & $151.31$ \\ 
		$0.3$ & $0.00761$ & $6.58$  & $4.52$ & $5.887$ & $14.554$ & $1.92$ & $11.146$ & $-1.09$ & $173.81$  \\ 
		$0.4$ & $0.00868$ & $4.18$  & $3.66$ & $6.678$ & $15.519$ & $1.92$ & $12.711$ & $-1.36$ & $185.53$  \\ 
		$0.5$ & $0.01121$ & $3.20$  & $3.13$ & $7.801$ & $18.070$ & $2.03$ & $15.848$ & $-1.60$ & $195.18$  \\ 
    \hline
  \end{tabular*}
\end{table*}

\subsection{Application to tetra-PEG hydrogels}
\label{sec:kamatta}

	Now, in order to validate our approach, we first consider the experimental data obtained from synthetically designed hydrogels with well controlled topology.
	As discussed in Refs.~\cite{kamata2014science,kamata2015advhmater}, the mechanical properties of tetra-PEG hydrogels are sensitive to changes of the polymeric chains about their affinity to water.
	It is shown in Ref.~\cite{kamata2014science} that, by increasing the molar fraction $r$ of the hydrophobic precursor molecules in the hydrogel, it is possible to control the swelling behaviour of such materials.

	Figure~\ref{fig:kamatta} includes the experimental data of Ref.~\cite{kamata2014science}, where the stress-strain curves were  determined for hydrogels synthetized with several values of $r$ at $T=37\,^{\text{o}}$C.
	As one may observe from Fig.~\ref{fig:kamatta}(a) as well as from Fig.~\ref{fig:kamatta}(b), where $\sigma(T,\gamma)/G(T)$ is plotted as a function of $\gamma/\gamma_s(T)$, our theoretical expression for the stress, Eq.~\eqref{eqn:sigma-gamma-q}, represents well the experimental data for the different values of $r$.
	Besides, Fig.~\ref{fig:kamatta}(c) indicates that the differential modulus $K(T,\gamma)$ numerically determined from the experimental data is also reasonably well described by Eq.~\eqref{eqn:differential-modulus}.
	In addition, from Fig.~\ref{fig:kamatta}(d), where it is plotted $K(T,\gamma)/G(T)$ against the reduced strain $\gamma/\gamma_{s}(T)$, 
one can clearly observe that the curves for $K(T,\gamma)$ present minima for finite strains.
	For clarity, we show in Figs.~\ref{fig:kamatta}(a),~(c), and~(d) only the experimental data obtained for two values of $r$, however, in Table~\ref{tab:kamatta_parameters} we include the values determined for the parameters so that the expressions~\eqref{eqn:sigma-gamma-q} and~\eqref{eqn:differential-modulus} can be consistently used to describe all the data extracted from Refs.~\cite{kamata2014science,kamata2015advhmater}.

	As indicated by the values of $G(T)$ and $K_{\min}(T)$ listed in Table~\ref{tab:kamatta_parameters}, the softening-induced behaviour becomes more pronounced as the ratio $r$ decreases and the energy difference $\Delta E$ increases.
	This is more evident, for instance, when the sample of PEG hydrogel with $r=0.2$ (green squares) is compared to the one with $r = 0.5$ (purple circles). 
	In fact, that result corroborates the behaviour somewhat expected from expression~\eqref{eqn:K-min}, 
since $K_{\min}(T)/G(T)$ is a decreasing function of $\Delta E$.

	The values presented in Table~\ref{tab:kamatta_parameters} also indicate that, as the ratio $r$ increases, the value of the maximum strain supported by the hydrogel $\gamma_s(T)$ also increases.
	This behaviour can be explained through Eq.~\eqref{eqn:gammas-q}, even though it is not apparent from theory how the parameters $n_e$,  $q_{\ell}^{\,}$, and $\Delta E$ should change as functions of $r$.
	One can empirically observe that $q_{\ell}$ increases with increasing values of $\Delta E$, with both quantities becoming larger as the ratio $r$ decreases.
	Consequently, one can infer from Eq.~\eqref{eqn:ell_0} that, by decreasing the fraction of hydrophobic chains in the network ({\it i.e.}, decreasing $r$), the natural end-to-end distance of the cross-links $\ell_{0}^{\,}(T)$ tends to increase.
	Hence, the increasing of $r$ reflects on reducing the values of $\gamma_s(T)$, that is, $(d \gamma_{s}(T)/d \Delta E) \propto -(1/\ell_0^2) (d \ell_{0}^{\,}/d \Delta E)$, since one has that $\gamma_s =(\ell_c - \ell_0^{\,})/\ell_0^{\,}$.
	This result corroborates the fact the smaller values for $\ell_0(T)$ are the source of large swelling processes observed for the PEG hydrogels with higher $r$ ratios described in Refs.~\cite{kamata2014science,kamata2015advhmater}.
	In fact, this behaviour is mentioned in Ref.~\cite{kamata2014science} as a definition, but here one sees that our model allows a quantitative discussion.

\subsubsection*{3.2.1~Analysis of the negative contribution}

	Since the material discussed here is a PEG hydrogel like the one studied in Refs.~\cite{yoshikawa2021prx,sakumichi2021polymj}, the existence of a negative energy-related contribution $G_E(T)$ to the elastic modulus is somewhat expected~\cite{duarte2023epje}.
	Indeed, as the values presented in Table~\ref{tab:kamatta_parameters} indicate, the value of the ratio $G_E(T)/G(T)$ increases in absolute value as the ratio $r$ of hydrophobic precursor molecules in the gel network increases.
	Besides, one finds from Eq.~\eqref{eqn:T_0*} that the values determined for the characteristic temperature $T_0^*$ are all positive, with higher values of $r$ leading to higher temperatures.
	As discussed before, Eq.~\eqref{eqn:G_E-G-w} confirms $T_0^* > 0$ as is an indicator of $G_E(T) < 0$.
	Interestingly, even though $\Delta E$ decreases with increasing values of $r$, the parameter $q_{\ell}^{\,}$ decreases faster, making $T_0^*$ an effectively increasing function of $r$.

\subsubsection*{3.2.2~Hardening effect}

	As already discussed from the data shown in Fig.~\ref{fig:collagen_intro}, 
the description of the hardening effect is a limitation of our model,
particularly when what is being observed in experiments is the rupture of the internal structures of the gel network after being subjected to a certain deformation.
	Yet, one can interpret $\gamma_s(T)$ as the maximum strain (on average) which the material may support before rupture, and that is the regime we are interested to describe here.

	Besides, it is worth noting that, while presenting the differential modulus, many studies in the literature~\cite{bertula2012acsmacrolett,licup2015pnas,wang2020angewchem,martikainen2020macromol,dealmeida2019natcommun} include plots of $K$ as a function of the stress $\sigma$ instead of the strain $\gamma$.
	In order to investigate how those physical quantities are related in our model, we consider Eq.~\eqref{eqn:sigma-gamma-q}, which can be rearranged as
\begin{equation}
\dfrac{1}{[1-\gamma/\gamma_s(T)][1+\gamma/\gamma_b(T)]} 
=\dfrac{\cosh^2\left[\left(\sigma/\bar{n}-\Delta E\right)/2k_BT\right]}{\cosh^2\left(\Delta E/2k_BT\right)}~,
\end{equation}
so that the differential modulus given by Eq.~\eqref{eqn:differential-modulus} can be rewritten in terms of the stress $\sigma$ as
\begin{equation}
K(T,\sigma) =K_\textrm{min}(T)\cosh^2\left(\dfrac{\sigma-\sigma_{\min}}{2\bar{n}k_BT}\right)~.
\label{eqn:diff-modulus-sigma}
\end{equation}
	Accordingly, the above expression indicates that the differential modulus $K(T,\sigma)$ will reach its minimum value $K_{\min}(T)$ when $\sigma$ is equal to the stress $\sigma_{\min}$ given by Eq.\eqref{eqn:sig_min}.

\begin{figure}[!t]
	\centering
	\resizebox{0.44\textwidth}{!}{\includegraphics{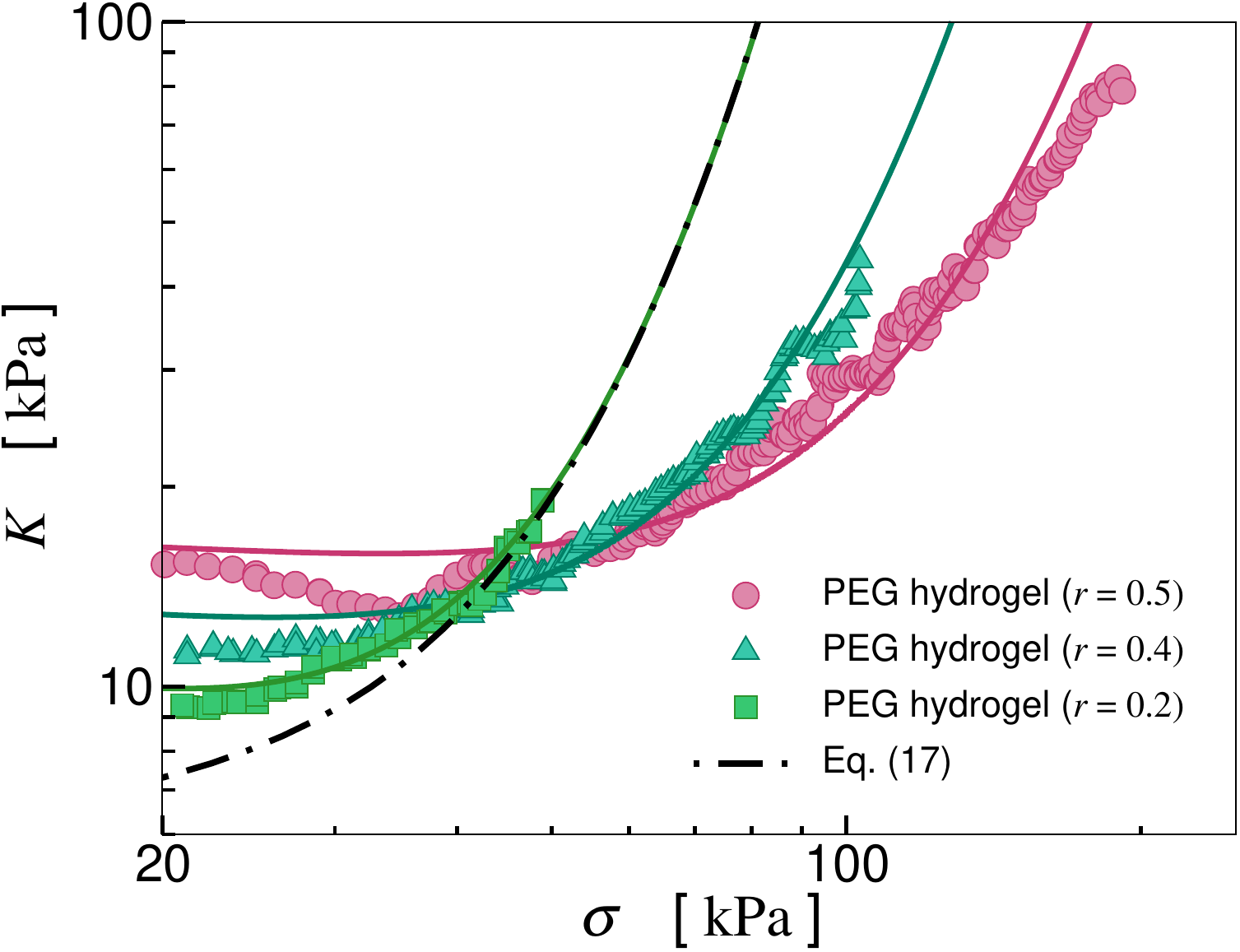}}
	\caption{Hardening effect observed from the differential modulus $K(\sigma,T)$ as a function of the stress $\sigma$.
	Filled symbols correspond to the differential modulus computed from the experimental data extracted from Ref.~\cite{kamata2014science} for PEG hydrogels with different $r$; continuous lines correspond to Eq.~\eqref{eqn:diff-modulus-sigma} evaluated using the parameters of Table~\eqref{tab:kamatta_parameters}.
	The exponential behaviour observed for large stresses is also described by the asymptotic expression~\eqref{eqn:diff-modulus-assynt}, which is included only for the sample with $r=0.2$ (dash-dotted line).
	}
	\label{fig:kamatta2}
\end{figure}

	In order to illustrate the behaviour of Eq.~\eqref{eqn:diff-modulus-sigma}, we include in Fig.~\ref{fig:kamatta2} the differential modulus $K$ as a function of the stress $\sigma$ evaluated from the experimental data obtained for PEG hydrogels of Ref.~\cite{kamata2014science} (which is also presented in Fig.~\ref{fig:kamatta}).
	Clearly, expression~\eqref{eqn:diff-modulus-sigma} describes reasonably well the experimental data, as can be verified from the continuous lines determined for the hydrogels prepared with $r =0.2$, $0.4$, and $0.5$.

	Now, by assuming relatively large values for the stress, {\it i.e.}, $\sigma\gg \sigma_{\min}+2\bar{n}k_BT$, one has that the asymptotic behaviour of the differential modulus given by Eq.~\eqref{eqn:diff-modulus-sigma} can be written as
\begin{equation}
K(T,\sigma) \approx \frac{1}{4}K_\textrm{min}(T)\left[2+e^{(\sigma-\sigma_\textrm{min})/\bar{n}k_BT}\right]~.
\label{eqn:diff-modulus-assynt}
\end{equation}
	As illustrated in Fig.~\ref{fig:kamatta2} by the curve determined for the hydrogel sample with $r=0.2$ ({\it i.e.}, dash-dotted line), the above asymptotic expression is suitable for describing $K(T,\sigma)$ for large values of stress.
	Interestingly, a similar asymptotic exponential behaviour, {\it i.e.}, $K\sim e^{c\sigma}$, with $c$ being a positive real constant, is also observed to occur for a model of rigid filaments connected by flexible cross-links, which is described by the free-energy density of the WLC model~\cite{heidenmann2015softmatter}.

\begin{figure*}[!t]
	\centering
	\resizebox{0.74\textwidth}{!}{\includegraphics{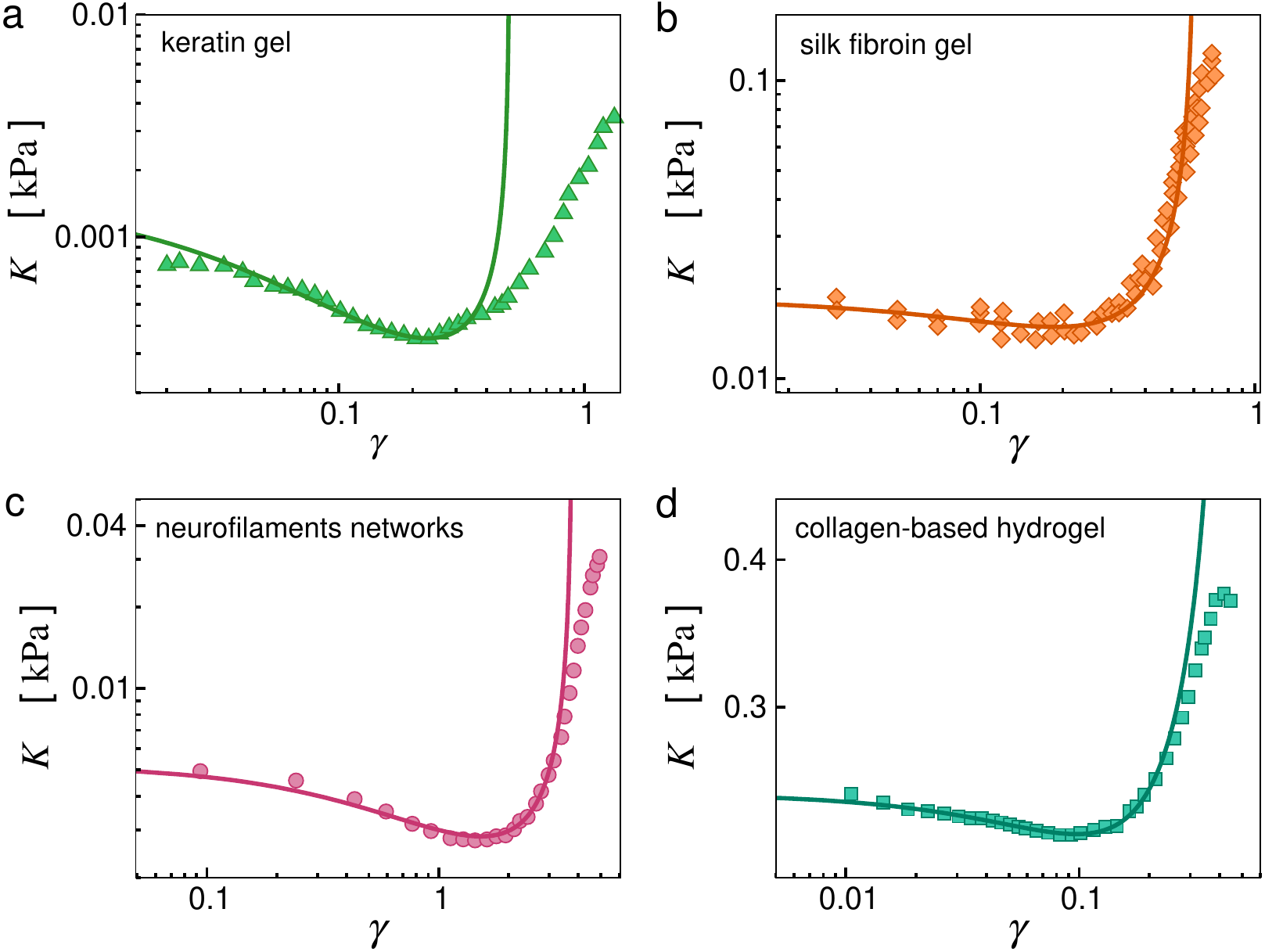}}
	\caption{Differential modulus $K(T,\gamma)$ given by Eq.~\eqref{eqn:differential-modulus} (continuous lines) for
		(a)~keratin gel at $25^{\circ}$C with $n_e = 1.05\times 10^{-5}\,$nm$^{-3}$, $q_{\ell} = -1.55$ and $\Delta E = 10.9\,$pN.nm; 
		(b)~silk fibroin gel at $25^{\circ}$C with $n_e = 4.3\times 10^{-6}\,$nm$^{-3}$, $q_{\ell} = -2.1$ and $\Delta E = 3.9\,$pN.nm;
		(c)~neurofilament network at $37^{\circ}$C with $n_e = 1.67\times 10^{-6}\,$nm$^{-3}$, $q_{\ell} = -18$ and $\Delta E = 7\,$pN.nm; 
		and 
		(d)~collagen-based hydrogel at $37^{\circ}$C with $n_e = 6.0\times 10^{-6}\,$nm$^{-3}$, 
$q_{\ell} = -1.95$ and $\Delta E = 2.33\,$pN.nm.
		Filled symbols denote the experimental data extracted from 
Refs.~\cite{valero2018plosone,storm2005nature,elbalasy2022polymers,oztoprak2017intjbiomol}.
	}
	\label{fig:biological_networks}
\end{figure*}

\subsection{Application to biopolymer-based hydrogels}
\label{sec:biologycal}

	Next, we consider the experimental data obtained for some biopolymer-based hydrogels studied in the literature in order to illustrate what other systems can be described by our model.
	First, it is worth exemplifying how one could relate the three parameters of the model ({\it i.e.}, $n_e$, $q_{\ell}$, and $\Delta E$), to match the experimental data by using the above theoretical results.
	In particular, one can use the differential modulus $K$ to graphically estimate its minimum value, $K_\textrm{min}(T)$, and also the elastic modulus, which is given either by $G(T) = \lim_{\gamma \rightarrow 0}K(T,\gamma)$ or by $G(T)= \lim_{\sigma \rightarrow 0}K(T,\sigma)$.
	Thus, from Eq.~\eqref{eqn:K-min}, one has that the energy difference is given by $\Delta E = 2k_BT\sech^{-1}(\xi)$, with $\xi=\sqrt{K_\textrm{min}(T)/G(T)}$.
	In addition, by estimating the value $\gamma_{\min}(T)$, one can use Eq.~\eqref{eqn:gamma-min} in a consistent manner to determine the maximum value of the strain, that is, $\gamma_{s}(T)=2\gamma_\textrm{min}(T)/(1-e^{-\Delta E/k_BT})$.
	Then, by using Eq.~\eqref{eqn:gammas-q}, one can estimate the ratio $q_{\ell} = [\gamma_{s}(T)+1]/[\gamma_{s}(T)e^{- \Delta E/k_BT}-1]$. 
	Finally, Eq.~\eqref{eqn:wT-q} can be used to determine the factor $\phi(T)$, so that the number density $n_e$ is obtained from Eq.~\eqref{eqn:elastic-modulus-w}, that is, $n_e = G(T)/k_BT\phi(T)$.

	Figure~\ref{fig:biological_networks} shows the experimental data obtained for (a)~keratin gel~\cite{elbalasy2022polymers}, (b)~silk fibroin gel~\cite{oztoprak2017intjbiomol}, (c)~neurofilament-based hydrogel~\cite{storm2005nature}, and (d)~collagen-based hydrogel~\cite{valero2018plosone}.
	By considering the above procedure, we found that the experimental data can be correctly described only if one assumes positive values for the energy difference, {\it i.e.}, $\Delta E > 0$, for all the biopolymer-based hydrogels considered. 
	Even so, differently from what is observed for the PEG hydrogels discussed in Sec.~\ref{sec:kamatta}, our analyses yielded only negative values for $q_{\ell}$ for the hydrogels displayed in Fig.~\ref{fig:biological_networks}.
	Consequently, since this ratio is defined as $q_{\ell}=-\ell_s/\ell_b$, both $\ell_s$ and $\ell_b$ must be positive parameters, so that the 
contributions of both end-to-end displacements are supposed to be in the same direction~\cite{duarte2023epje}.

	In addition, because $q_{\ell}< -1$, one cannot determine a threshold temperature as 
$T_0^*$ given by Eq.~\eqref{eqn:T_0*}, so these hydrogels will not present any positive temperature $T$ where the elastic moduli $G(T)$ reaches zero~\cite{duarte2023epje}, as in the case where $q_{\ell}>1$.
	Even so, since $q_{\ell}< -1$, the energy-related contribution $G_E(T)$ given by Eq.~\eqref{eqn:G_E-G-w} can be either positive or negative, depending on the temperature. 
	By setting $q_{\ell} = -|q_{\ell}|$ in Eq.~\eqref{eqn:G_E-G-w} one finds that, if $T>T^\dagger_0$, then $G_E(T)<0$, while for $T<T^\dagger_0$, it yields $G_E(T)>0$, with such temperature given by
\begin{equation}
T^\dagger_0 = \dfrac{2\Delta E}{k_B\ln(-q_{\ell}^{\,})}
\label{eqn:Tc}
\end{equation}
	Interestingly, this last result suggests a way of testing our model experimentally, since the sign of $G_E(T)$ would change at the predicted temperature $T^\dagger_0$ given by Eq.~\eqref{eqn:Tc}.
	The problem is that, in some cases, $T^\dagger_0$ might be estimated as a very high temperature, so that the samples would be degraded when subjected to the hypothetical experiment.
	That is the case of the keratin ($T_0^\dagger\approx 3331^{\circ}$C), silk fibroin ($T_0^\dagger\approx 489^{\circ}$C) and, possibly, neurofilaments ($T_0^\dagger\approx 78^{\circ}$C) gels, which are presented in Fig.~\ref{fig:biological_networks}.
	Yet, we believe that it is possible to find hydrogels from which the experiments can be performed without either damaging the gel network or boiling the solvent.
	One example is the collagen-based hydrogel that provided the results displayed in Fig.~\ref{fig:biological_networks}(d) (which is the same data presented in Fig.~\ref{fig:collagen_intro}), for which the temperature is found to be $T_0^\dagger\approx 17^{\circ}$C.

	Here, it is worth noting that, because the experimental data presented in Fig.~\ref{fig:biological_networks} were obtained either at room or physiological temperatures, the collagen-based hydrogel is the only one for which the energy-related contribution to the elastic modulus is negative, {\it i.e.}, $G_E(T)/G(T) =-0.008$.
	For the keratin hydrogel one has the most significant positive energy-related contribution, being over twice its total elastic modulus, {\it i.e.}, $G_E(T)/G(T) = 2.21$, while for silk fibroin and neurofilaments gels we found $G_E(T)/G(T) = 0.233$ and $G_E(T)/G(T) = 0.096$, respectively.

	In addition, we found for the biopolymer-based hydrogels that
$\gamma_s=0.5$ (keratin), 
$\gamma_s=0.61$ (silk fibroin), 
$\gamma_s=3.77$ (neurofilaments), 
and $\gamma_s=0.45$ (collagen). 
	Although Eq.~\ref{eqn:differential-modulus} is not able to describe the experimental data at large strains (possible due to the breakage or to the heterogeneous character~\cite{rizzi2020jrheol} of the internal structures of the networks), these values present the opposite trend observed for tetra-PEG hydrogels (see Table~\ref{tab:kamatta_parameters}), {\it i.e.}, the higher the value of $|q_{\ell}|$, the larger the value of $\gamma_s$.

\section{Concluding remarks}

	In this work we have investigated the possibility that the coarse-grained model introduced in Ref.~\cite{duarte2023epje} could be used to describe the strain-induced softening behaviour observed in many hydrogels.
	Despite its simplicity, the two-state model led to the stress-strain relationship given by Eq.~\eqref{eqn:sigma-gamma-q}, from where we were able to determine exact expressions for the differential modulus $K$ as a function of both the strain, Eq.~\eqref{eqn:differential-modulus}, and the stress, Eq.~\eqref{eqn:diff-modulus-sigma}.

	We showed that those expressions can be used to fully describe the mechanical response and the thermal behaviour of several hydrogels with just three physically meaningful parameters, {\it i.e.}, the number density of chains which connect cross-links, $n_e$, the ratio between end-to-end distances, $q_{\ell}^{\,}= - \ell_s/\ell_b$, and the energy difference between the two states, $\Delta E=E_s - E_b$.

	The expressions obtained for the differential modulus indicate that, if $\Delta E = 0$, the model describes purely entropic hydrogels which do not display the strain-induced softening behaviour, so only the hardening effect is observed.
	When $\Delta E>0$, the model predicts that both the strain-induced softening and the hardening behaviours will occur.
	In that case, our results indicate that the softening behaviour can be observed for hydrogels with both positive and negative values of $q_{\ell}^{\,}$, as exemplified by the data presented in Figs.~\ref{fig:kamatta} and \ref{fig:kamatta2} for PEG hydrogels, where $q_{\ell}^{\,}>1$, and in Fig.~\ref{fig:biological_networks} for biopolymer-based gels, where $q_{\ell}^{\,}<-1$.

	By finding that the energy difference $\Delta E$ is positive, we were able to infer the importance of effective interactions between the chains in the gel network and their neighbouring solvent molecules.
	This energy difference not only leads to the presence of the energy-related contribution $G_E(T)$, as discussed in Ref.~\cite{duarte2023epje}, but it also induces, as we show here, a solvent-related strain-induced softening behaviour.
	Although solvent-induced effects have been previously considered in Ref.~\cite{yoshikawa2021prx}, our results revealed that their phenomenological model, like those based on the WLC model~\cite{palmer2008acta}, is not capable 
of displaying a softening behaviour.

	It is worth emphasizing that, in contrast to purely entropic~\cite{flory1953book,james1953jcp,flory1977jcp} and to purely enthalpic~\cite{bouzid2018langmuir,heidenmann2015softmatter,huisman2010pre,huisman2011prl} approaches, our statistical mechanical model takes into account both contributions.
	Even so, our model is limited to situations where there is no rupture or yielding of the internal structure of the gel network, since it obviously involves, as already mentioned, a different solvent-induced mechanism for the softening and hardening behaviours.
	As discussed along the manuscript, the hardening behaviour that is observed at large strains from our model corresponds to an exponential behaviour, {\it i.e.}, Eq.~\eqref{eqn:diff-modulus-assynt}, just like the asymptotic behaviour found in Ref.~\cite{heidenmann2015softmatter} from a completely different approach.

	Finally, we note that the theoretical basis of our model may serve as a first approximation for modelling approaches that aim to describe more complex hydrogels related to biomedical applications~\cite{guidetti2022aprev,greco2023biophysrev,nicolle2017jmechbiomed}.




The authors thank the funding agencies FAPEMIG (APQ-02783-18), CNPq (312999/2021-6), and CAPES.







\vspace{-0.3cm}

\section*{References}

\vspace{-0.65cm}


%

\end{document}